\documentclass[11pt,twoside]{article}

\usepackage{asp2014}

\aspSuppressVolSlug
\resetcounters

\begin{document}

\title{RCSEDv2: analytic approximations of k-corrections for galaxies out to redshift $z=1$}

\author{Anastasia~Kasparova,$^{1}$ 
Igor~Chilingarian,$^{2,1}$ 
Sviatoslav~Borisov,$^{3,1}$
Vladimir~Goradzhanov,$^{1,4}$
Kirill~Grishin,$^{5,1}$
Ivan~Katkov,$^{6,1}$ 
Vladislav~Klochkov,$^{1,4}$
Evgenii~Rubtsov,$^{1,4}$ and
Victoria~Toptun$^{1,4}$}
\affil{$^1$Sternberg Astronomical Institute, Moscow State University, Moscow, Russia; \email{anastasya.kasparova@gmail.com}}
\affil{$^2$Center for Astrophysics -- Harvard and Smithsonian, Cambridge, MA, USA}
\affil{$^3$Department of Astronomy, University of Geneva, Versoix, Switzerland}
\affil{$^4$Department of Physics, Moscow State University, Moscow, Russia}
\affil{$^5$Universite de Paris, CNRS, Astroparticule et Cosmologie, F-75013 Paris, France}
\affil{$^6$New York University Abu Dhabi, Abu Dhabi, UAE}

\paperauthor{Anastasia Kasparova}{anastasya.kasparova@gmail.com}{0000-0002-1091-5146}{Sternberg Astronomical Institute, Lomonosov Moscow State University}{}{Moscow}{}{119234}{Russia}
\paperauthor{Igor~Chilingarian}{igor.chilingarian@cfa.harvard.edu}{ORCID}{Center for Astrophysics -- Harvard and Smithsonian / Sternberg Astronomical Institute}{}{Cambridge}{MA}{02138}{USA}
\paperauthor{Sviatoslav Borisov}{sb.borisov@voxastro.org}{0000-0002-2516-9000}{University of Geneva}{Department of Astronomy}{Geneva}{}{}{Switzerland}
\paperauthor{Vladimir Goradzhanov}{goradzhanov.vs17@physics.msu.ru}{0000-0002-2550-2520}{Sternberg Astronomical Institute, Lomonosov Moscow State University}{}{Moscow}{}{119234}{Russia}
\paperauthor{Kirill Grishin}{kirillg6@gmail.com}{0000-0003-3255-7340}{Sternberg Astronomical Institute, Lomonosov Moscow State University}{}{Moscow}{}{119234}{Russia}
\paperauthor{Ivan Katkov}{katkov.ivan@gmail.com}{0000-0002-6425-6879}{NYU Abu Dhabi}{Center for Astro, Particle, and Planetary Physics}{Abu Dhabi}{}{129188}{UAE}
\paperauthor{Vladislav Klochkov}{vladislavk4481@gmail.com}{0000-0003-3095-8933}{M.V. Lomonosov Moscow State University}{Department of Physics}{Moscow}{}{119991}{Russia}
\paperauthor{Evgenii Rubtsov}{rubtsov602@gmail.com}{0000-0001-8427-0240}{Sternberg Astronomical Institute, Lomonosov Moscow State University}{}{Moscow}{}{119234}{Russia}
\paperauthor{Victoria Toptun}{victoria.toptun@voxastro.org}{0000-0003-3599-3877}{Sternberg Astronomical Institute, Lomonosov Moscow State University}{}{Moscow}{}{119234}{Russia}

\begin{abstract}
To compare photometric properties of galaxies at different redshifts, we need to correct  fluxes for the change of effective rest-frame wavelengths of filter bandpasses, called $k$-corrections. At redshifts $z>0.3$, the wavelength shift becomes so large that typical broadband photometric bands shift into the neighboring rest frame band. At $z=0.6-0.8$ the shift reaches two or even three bands. Therefore, we need perform $k$-corrections from one observed bandpass to another. Here we expand the methodology proposed by \citet{2010MNRAS.405.1409C} and fit cross-band $k$-corrections by smooth low-order polynomial functions of one observed color and a redshift -- this approach but without cross-band is implemented as standard functions in {\sc topcat}, which can be used for galaxies at $z<0.5$. We also computed analytic approximations for WISE bands, which were not available in the past. We now have a complete set of $k$-corrections coefficients, which allow us to process photometric measurements for galaxies out to redshift $z=1$. We calculated standard and cross-band $k$-corrections for about 4 million galaxies in second Reference Catalog of Spectral Energy Distributions (RCSEDv2) of galaxies and we showed that, in cases of widely used UV, optical and near-infrared filters, our analytic approximations work very well and can be used for extragalactic data from future wide-field surveys.
\end{abstract}

\section{Motivation}

To study galaxy evolution we need compare photometric properties of galaxies at different redshifts by correcting fluxes for the changes of effective rest-frame wavelengths of filter bandpasses, called $k$-corrections. At redshifts exceeding $z=0.3$, the wavelength shift becomes so large that typical broad photometric bands shift into the neighboring rest-frame bands (e.g. an observed SDSS $r$ into a rest-frame SDSS $g$). At redshifts approaching $z=0.7$ the shift reaches two bands. Therefore, for practical purposes we need calculate cross-band $k$-corrections, which, when applied, will convert an observed magnitude in a given band to a rest-frame magnitude in another band. As it was demonstrated by \citet{2010MNRAS.405.1409C} and \citet{2012MNRAS.419.1727C}, $k$-corrections for non-active galaxies at $z<0.5$ can be precisely fitted by low-order two-dimensional polynomial functions of a redshift and a properly chosen observed color. This approach has a great advantage over widely used fitting of a broadband spectral energy distribution (SED) implemented e.g. in the {\sc kcorrect} package \citep{2007AJ....133..734B} because it requires only one observed color rather than a full SED. The coefficients from \citet{2012MNRAS.419.1727C} are implemented as kCorr\_*() functions in {\sc topcat} \citep{2005ASPC..347...29T} and were used to compute $k$-corrections in the RCSED project \citep{2017ApJS..228...14C}. Here we expand the low-order fitting approach to cross-band $k$-corrections, which one can use up-to a redshift $z=1$.

\begin{figure}
    \centering
    \includegraphics[width=0.59\hsize]{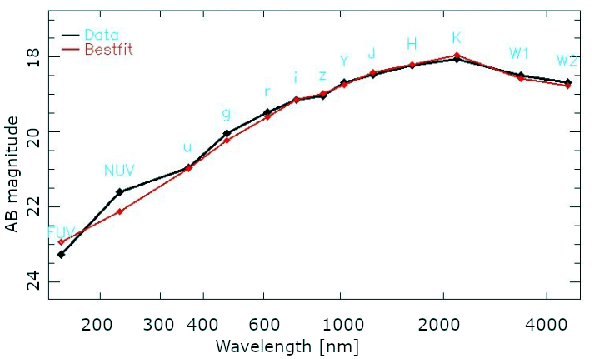}
    \includegraphics[width=0.39\hsize]{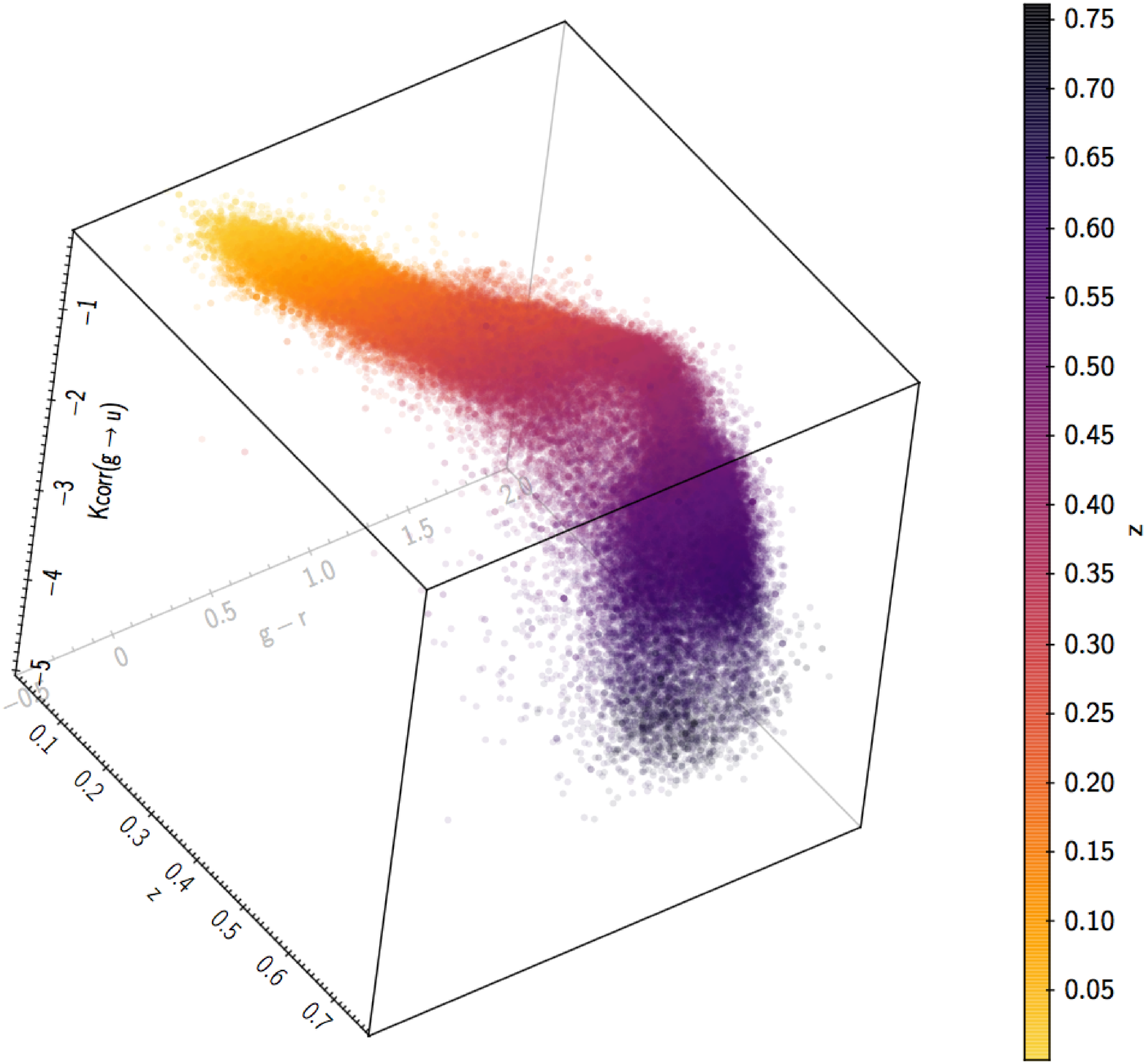}
    \caption{Left: example of a broadband SED fitting using a set of {\sc kcorrect} templates. Right: the 3D distribution of galaxies in the (z, $g-r$, $k_{corr}(g\rightarrow u)$ space with the redshift color-coded.}
    \label{ex_fig1}
\end{figure}

\begin{figure}
\includegraphics[width=0.49\hsize]{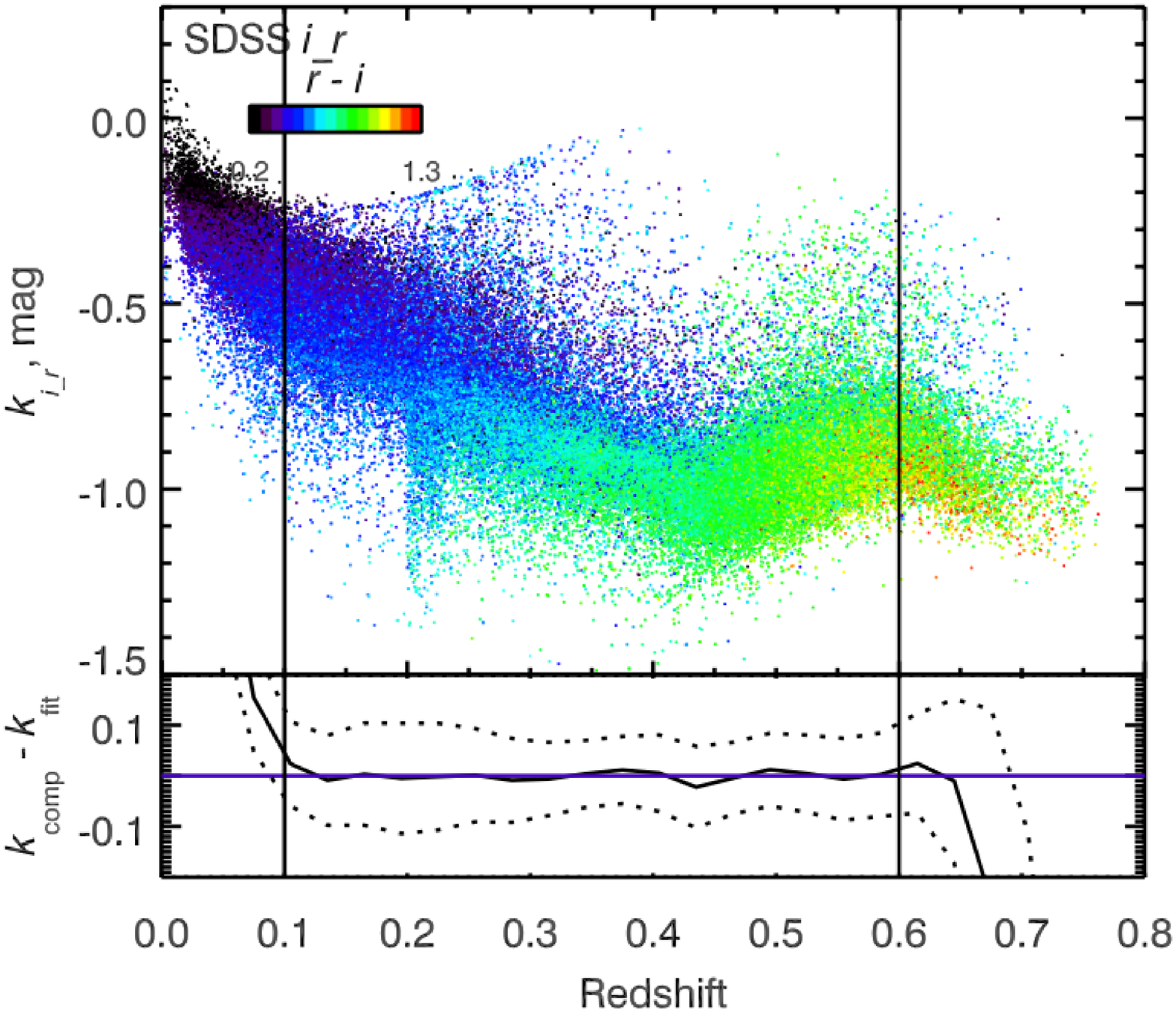}
\includegraphics[width=0.49\hsize]{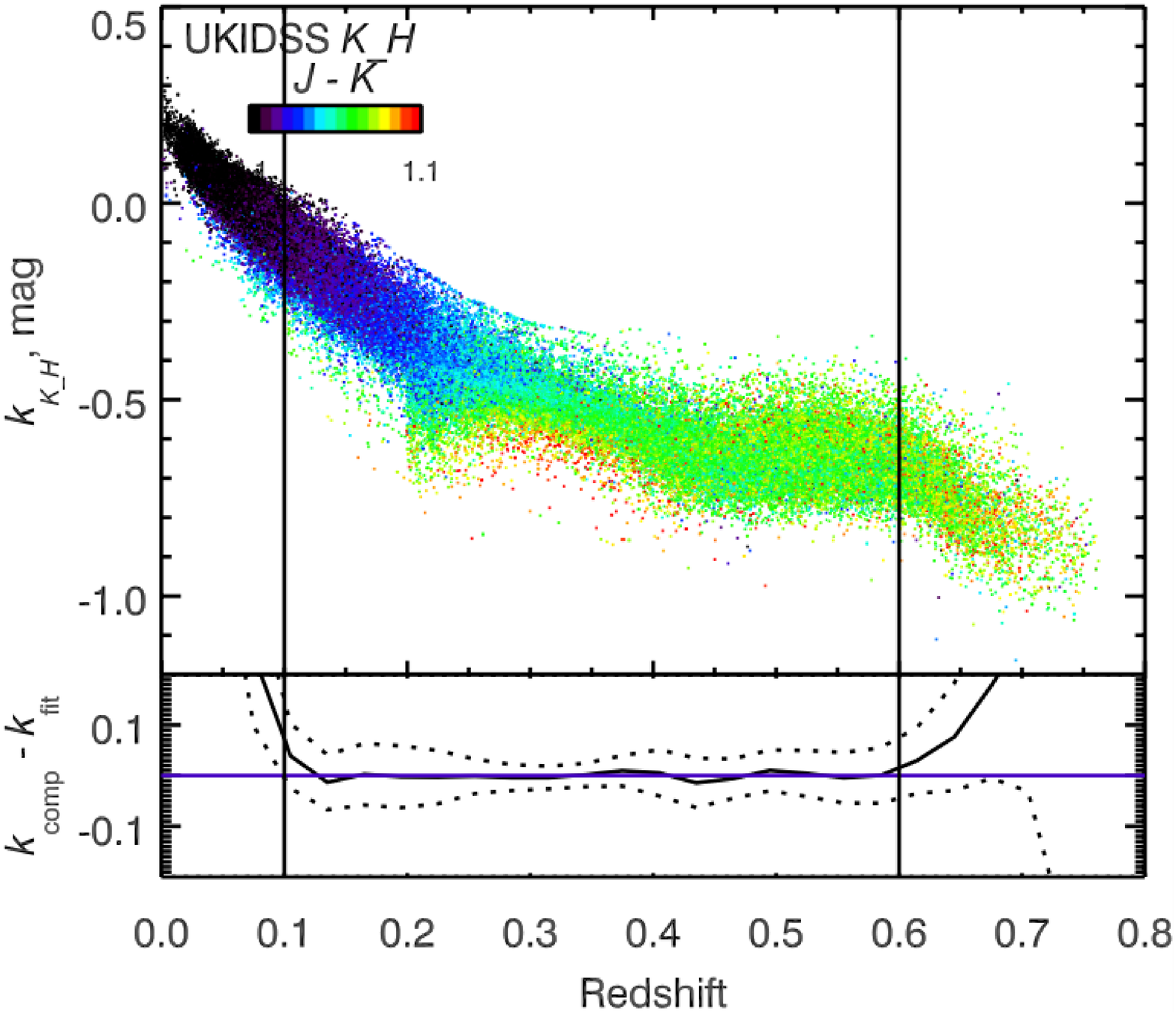}
\caption{The fitting examples of the cross-band K-corrections ($z=0.3$) in optical and IR bands.} \label{fig_fits}
\end{figure}

\section{Computing and fitting $k$-corrections}

Here we follow the algorithm described in \citet{2012MNRAS.419.1727C}: we built a sample of high-quality photometric measurements covering the desired redshift range and fitted them using a linear combination of galaxy templates (see Fig.~\ref{ex_fig1}, left) produced by the non-negative matrix inversion and used in the {\sc kcorrect} package \citep{2007AJ....133..734B}. We also took into account internal dust extinction inside galaxies by fitting the $e(B-V)$ value. 

Our input photometric dataset was extracted as a subset of the RCSEDv2 catalog with the most accurate photometric optical/IR data from the following surveys: GALEX (near-UV; \citealp{2005ApJ...619L...1M}), SDSS \citep{SDSS_DR7}, DESI Legacy Surveys \citep{2019AJ....157..168D}, Dark Energy Survey \citep{2021ApJS..255...20A}, UKIDSS \citep{2007MNRAS.379.1599L}, CatWISE \citep{2020ApJS..247...69E}. The datasets were retrieved using publicly available Virtual Observatory access services. Having the best-fitting results at hand we then computed restframe fluxes for the very same templates and derived $k$-corrections.

Then we explored the derived $k$-corrections in the ($z - color - k_{corr}$) 3D parameter space and selected the appropriate degrees of the 2D fitting polynomial. An example of a 3D distribution for $k_{corr}(g\rightarrow u)$ is shown in Fig.~\ref{ex_fig1} (right).

Then we fitted scattered datasets and derived the coefficients of the best-fitting surfaces. However, this approach being applied directly to intermediate-redshift data leads to poor quality of the solution because of a non-uniform redshift distribution of galaxies in the redshift--color parameter space: a densely populated area of the parameter space will dominate the $\chi^2$ value causing systematic offsets in the sparsely populated regions usually located close to the edges of the parameter space. To tackle this issue, instead of fitting full scattered datasets, we first applied a two-dimensional binning on a fine grid (e.g. 0.02 in redshift and 0.015--0.03~mag in color depending on the band) and used the median values of $k$-corrections inside each bin in the fitting procedure. We excluded from the fitting procedure the bins where a number of data points was below a certain threshold ($N=15$). We also slightly modified the polynomial fitting procedure compared to that used in \citet{2010MNRAS.405.1409C,2012MNRAS.419.1727C} because there we fixed a global offset (0th order coefficient) to zero using the fact that $k_{corr}(z=0)=0$ by definition. Here, because we compute cross-band $k$-corrections, this assumption is not valid and should not be used.

To evaluate the quality of analytic approximations, we then computed the best-fitting $k$-correction values for every individual galaxy and analyzed the fitting residuals. Fig.~\ref{fig_fits} presents examples of the computed $k$-corrections vs redshift with a color-coded observed color used to perform the fitting. Lower panels display the residuals between the analytic approximation and measured values (bold black line) and r.m.s. of the fitting residuals (dotted lines).


\section{Results and implications}

We computed polynomial approximations of cross-band $k$-corrections in UV, optical and IR broad-band filters as functions of redshift and observed colours for galaxies at redshifts $z = 0.3$, $z = 0.5$ and $z = 1.0$. We also computed approximations of WISE $W1$ and $W2$ $k$-corrections at low redshifts, which were not available in the past. The coefficients will be integrated into an upgraded version of the $k$-corrections calculator (\url{http://kcor.sai.msu.ru/}) and after a thorough testing and validation will be supplied to the developer of {\sc topcat} to be included in one of the next releases of the tool. 

We use these approximations in RCSEDv2 (\url{https://rcsed2.voxastro.org/}), the second version of RCSED that is substantially extended to higher redshifts and also contains WISE photometry for over 4 million galaxies. The crucial importance of accurate computation of $k$-corrections is illustrated by the fact that the measured width of the red sequence in the $M_r - (g-r)$ color space (and the width of the universal UV-optical color--color--magnitude relation in the red end; see \citealp{2012MNRAS.419.1727C}) is $0.045$~mag  and the intrinsic width is probably even narrower. This important property of passive galaxies is used to select rare compact elliptical \citep{2015Sci...348..418C} and low-mass post-starburst galaxies \citep{2021NatAs.tmp..208G} as outliers from the relations. Such selection techniques would fail if $k$-correction values are biased.

The next step is to compute analytic approximations of $k$-corrections for active galactic nuclei (AGN) and quasars. This task is, however, much more complex. An AGN host galaxy might have a substantial flux contribution from the accretion disk around the AGN to the total flux in the optical, UV, and mid-IR bands even for relatively low-mass central black holes if they happen to accrete close to the Eddington limit \citep{2018ApJ...863....1C}. This contribution is known to correlate with the H$\alpha$ (and H$\beta$) flux in the broad-line component, which can be measured directly from spectra. Therefore, the broad-line H$\alpha$ flux is an obvious choice as an additional fitting parameter.

\acknowledgements This project is supported by the RScF grant 19-12-00281 and the Interdisciplinary Scientific and Educational School of Moscow University ``Fundamental and Applied Space Research''.

\bibliography{X7-004.bib}  

\end{document}